\documentclass[aps,prd,floatfix,showpacs,10pt,twocolumn,nofootinbib]{revtex4-1}
\usepackage{amsmath}
\usepackage{amsfonts}
\usepackage{amssymb}
\usepackage{amsthm}
\usepackage{mathrsfs}
\usepackage{multirow}
\usepackage{color}
\usepackage{cancel}
\usepackage{array}
\usepackage{tikz}
\usetikzlibrary{calc}
\usetikzlibrary{decorations}
\usetikzlibrary{decorations.fractals}
\usetikzlibrary{decorations.pathreplacing}
\usetikzlibrary{decorations.pathmorphing}
\usetikzlibrary{decorations.markings}
\usetikzlibrary{positioning,matrix}
\usetikzlibrary{shapes.arrows}
\usetikzlibrary{shapes.symbols}
\usetikzlibrary{shapes.misc}
\usetikzlibrary{shapes.geometric}
\usetikzlibrary{through}
\usetikzlibrary{mindmap,backgrounds}
\usetikzlibrary{fit}
\usetikzlibrary{arrows,positioning}
\usetikzlibrary{external}
\usepackage[colorlinks, urlcolor=black, citecolor=blue, link
color=black]{hyperref}

\def\nn    {\nonumber}

\def\beq{\begin{equation}}
\def\eeq{\end{equation}}
\def\beqa{\begin{eqnarray}}
\def\eeqa{\end{eqnarray}}

\allowdisplaybreaks

\begin{document} 
\title{Unraveling the couplings of a Drell-Yan produced $Z'$ with heavy-flavor tagging}

\author{Wei-Shu Hou$^{1,2}$, Masaya Kohda$^1$, and Tanmoy Modak$^1$}
\affiliation{$^1$Department of Physics, National Taiwan University, Taipei 10617, Taiwan}
\affiliation{$^2$ARC CoEPP at the Terascale,
School of Physics, University of Melbourne,
Vic 3010, Australia}
\bigskip

\vspace{.2cm}

\begin{abstract}
Despite no new physics so far at the LHC, a $Z'$ boson with $m_{Z'} \sim 100$ GeV 
could still emerge via Drell-Yan (DY) production, $q \bar q \to Z' \to \mu^+ \mu^-$,
in the next few years.  
To unravel the nature of  the $Z'$ coupling, we utilize the $c$- and $b$-tagging 
algorithms developed by ATLAS and CMS to investigate 
$cg \to c Z'$  at 14 TeV LHC. 
While light-jet contamination can be eliminated,
mistagged $b$-jets cannot be rejected in any of 
the tagging schemes we adopt. 
On the other hand, for nonzero $bbZ'$ coupling,  
far superior $b$-tagging could discover the $bg \to b Z'$ process,
where again light-jet mistag can be ruled out, 
but mistagged $c$-jets cannot yet be excluded. 
Provided that DY production is discovered soon enough,
we find that a simultaneous search for $c g \to c Z'$ and $b g \to b Z'$ 
can conclusively discern the nature of $Z'$ couplings involved.
\end{abstract}

\maketitle

\section{Introduction}
\label{intro}

A $Z'$ boson a few hundred GeV in mass could still emerge 
via the Drell-Yan (DY) process, $q \bar q \to Z' \to \mu^+ \mu^-$, 
for $qqZ'$ couplings that are weaker than analogous Standard Model (SM) couplings. 
Recent searches~\cite{Aaboud:2017buh, CMS:2016abv}
set stringent bounds on the couplings of such a $Z'$ boson
to $u$, $d$ and $s$ quarks,
but the limits are much weaker for $c$ or $b$ quarks, 
hence discovery is possible within the next few years. 
One such scenario~\cite{Hou:2017ozb} involves a $Z'$ that couples to $c$ quarks,
leading to DY production $c \bar c \to Z' \to \mu^+ \mu^-$ at the LHC. 
The $c g \to c Z' \to c \mu^+ \mu^-$ process then offers a unique probe 
of the flavor structure of the $Z'$ coupling, 
if the $c$-jet flavor can be identified. 
Recent developments at ATLAS and CMS in 
$c$-tagging~\cite{atlasctag8tev,atlasctag,CMS:2016knj} algorithms
and excellent performance of $b$-tagging~\cite{atlasctag,Aad:2015ydr,CMS:2016kkf} 
offer such an opportunity.  
In this paper we discuss how these heavy flavor taggers
can probe the couplings of a $Z'$ after its discovery through the DY process.

We illustrate with the scenario of Ref.~\cite{Hou:2017ozb}, 
where a $Z'$ couples relatively weakly to charm quarks and predominantly to muons.
The DY process $pp \to Z'+X \to \mu^+ \mu^-+X$ ($X$ being inclusive activity) 
could emerge in the next few years, 
and a $ccZ'$ coupling would imply the $cg\to cZ'$ process.
We apply the $c$-tagging algorithms to investigate
the discovery potential of $pp\to c Z'+ X \to c \mu^+ \mu^- +X$ 
(denoted as $cZ'$ process, with the conjugate process implied) 
at $\sqrt{s}=14$ TeV LHC. 

The $c$-tagging algorithms of ATLAS~\cite{atlasctag8tev,atlasctag} 
and CMS~\cite{CMS:2016knj} discriminate $c$-jets from light-jets 
(jets originating from $u$, $d$, $s$ and gluon) 
at the expense of $c$-tag efficiency,
while misidentification (or mistag) rate of $b$-jets as $c$-jets are relatively sizable. 
If the $Z'$ couples to light $ q= u$, $d$, $s$ quarks, 
a potential $cZ'$ signal may arise from mistag
(denoted as fake $cZ'$). As the $qqZ'$ coupling is constrained
by search for heavy resonance in DY process~\cite{Aaboud:2017buh}, 
our analysis shows that in certain $c$-tagging schemes one can 
completely rule out the possibility of fake $cZ'$ from light-jets. 
But these tagging schemes fail to rule out the 
possibility of fake $cZ'$ from mistagged $b$-jet.

In case the $Z'$ couples instead to $b$ quarks ($bbZ'$ coupling), 
$pp\to b Z'+ X \to b \mu^+ \mu^- +X$ ($bZ'$ process) 
would emerge after the discovery in DY. 
This process could be observed by the well developed $b$-tagging 
algorithms~\cite{atlasctag,Aad:2015ydr,CMS:2016kkf},
which provide excellent discrimination against light- and $c$-jets
while maintaining high $b$-tagging efficiency. 
We find the current limit on $ccZ'$ coupling 
allows for fake $bZ'$ discovery at LHC
due to mistag of $c$-jet as $b$-jet. 
However, this fake $bZ'$ process at LHC could be ruled out
if $\sim 250$ fb$^{-1}$ data is collected. 
We find that, if a $Z'$ is discovered via the DY process in the next few years, 
combining the $cZ'$ and $bZ'$ signatures together with current limits
from heavy resonance searches, 
one can conclusively infer the nature of $Z'$ couplings.

We finally consider a case where both $bbZ'$ and $ccZ'$ couplings are nonzero 
and study DY, $cZ'$ and $bZ'$ processes for a representative $Z'$ mass.
We find that the coupling structure of such a scenario can also be disentangled, 
if combined with the current limit from heavy resonance search in DY process.

The paper is organized as follows. 
In Sec.~\ref{dy}, we analyze the discovery potential of 
the DY process due to $qqZ'$ couplings. 
In Sec.~\ref{czprime}, we apply different $c$-tagging algorithms  
for the discovery potential of the $cZ'$ process and discuss fake sources. 
Sec.~\ref{bzprime} is dedicated to the $bZ'$ process, and 
on disentangling the $Z'$ coupling structure by combining with 
the results of Sec.~\ref{czprime}. 
The scenario for having both $ccZ'$ and $bbZ'$ couplings is analyzed 
in Sec.~\ref{bothnonzero}, and we summarize in Sec.~\ref{summary}.
The analysis for the DY process is detailed in Appendix~\ref{dimuon}, 
while normalized kinematic distributions for the signal and backgrounds 
of the $cZ'$ process are provided in Appendix~\ref{dist}.

\begin{figure*}[htbp!]
\centering
\includegraphics[width=.35 \textwidth]{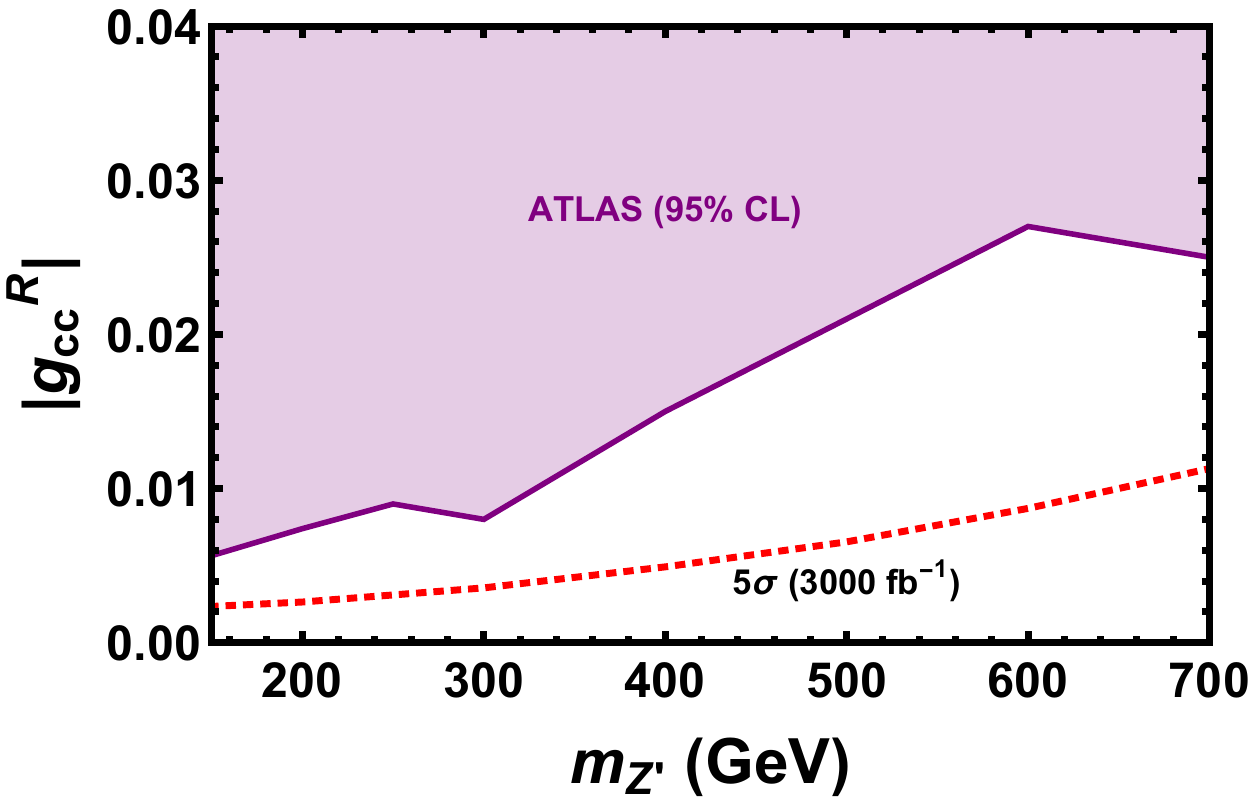}
\includegraphics[width=.35 \textwidth]{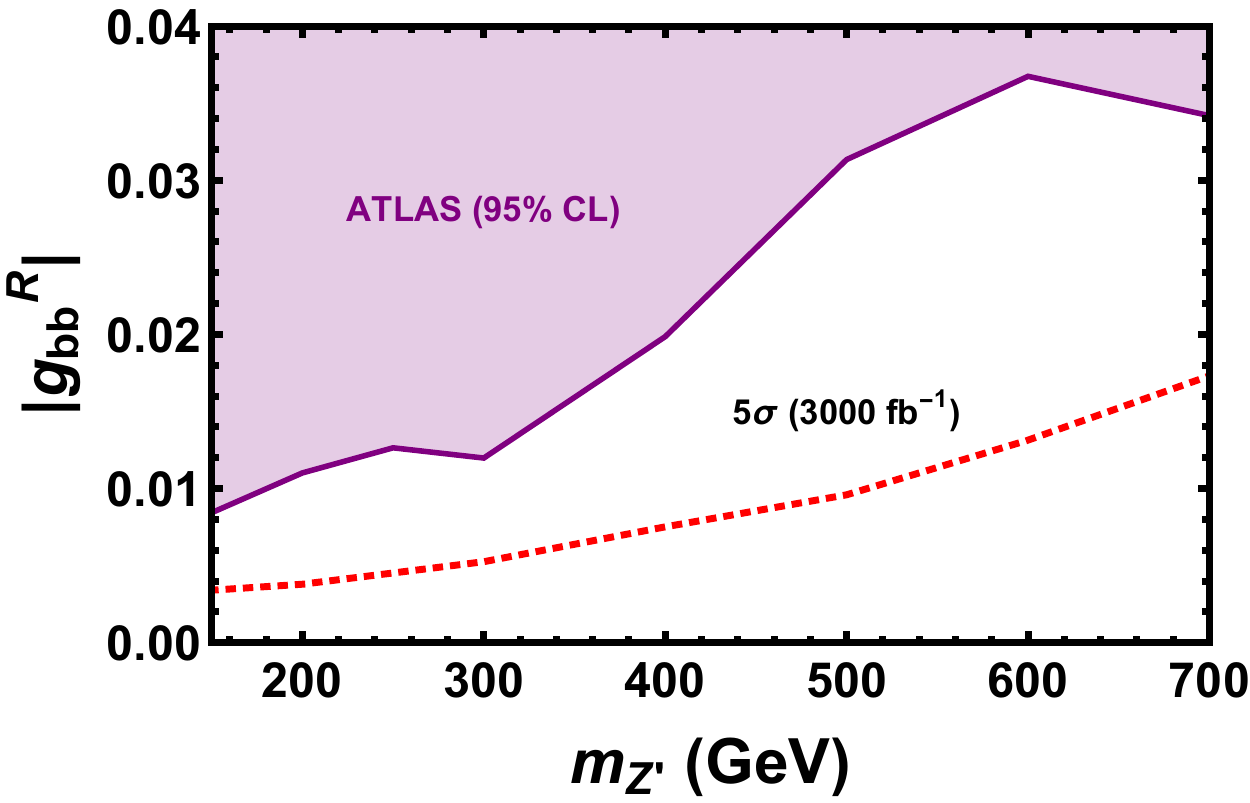}
\caption{
 The $5\sigma$ discovery reach of the DY process $pp\to Z' + X \to \mu^+\mu^- + X$ 
 at 14 TeV LHC with 3000 fb$^{-1}$ data,
 initiated by $ccZ'$ (left) and $bbZ'$ (right) couplings.
 The purple shaded regions are the 95\% CL upper limits
 extracted from Ref.~\cite{Aaboud:2017buh}. 
}
\label{dyprocess}
\end{figure*}
%

\section{The Drell-Yan process}
\label{dy}

We take the following effective couplings,
\begin{align}
 \mathcal{L} \supset & 
-g'\bigg( \bar{\mu}\gamma_\alpha \mu+\bar{\nu}_{\mu L} \gamma_\alpha \nu_{\mu L}
             -\bar{\tau}\gamma_\alpha \tau-\bar{\nu}_{\tau L} \gamma_\alpha \nu_{\tau L}
       \bigg) Z'^{\alpha} \nn\\
 & -\sum_{q=u,d,s}^{c,b} g^R_{qq} \, \bar{q}_R\gamma_\alpha q_R\, Z'^\alpha, 
\label{lepton}
\end{align}
where $g'$ is the coupling of  $Z'$ to the muon, tauon and their neutrinos,
and $g^R_{qq}$ is the right-handed (RH) $qqZ'$ coupling
(induced by some underlying heavy particles~\cite{Fuyuto:2015gmk}). 
The context is the effective model based on the
gauged $L_\mu-L_\tau$~\cite{He:1990pn, Foot:1990mn} symmetry, 
as discussed in Refs.~\cite{Fuyuto:2015gmk, Altmannshofer:2014cfa}. 
For simplicity and to be more general, we set all flavor violating
couplings to zero and assume $g^R_{qq}$ to be real. 
The coupling $g'$ is taken to be much larger than the coupling $g^R_{qq}$, 
hence the $Z'$ couples more weakly to quarks, 
and its decay branching ratios can be approximated as:
\begin{align}
&\mathcal{B}(Z'\to\mu^+\mu^-) \simeq \mathcal{B}(Z'\to\tau^+\tau^-) \simeq \mathcal{B}(Z'\to\nu\bar\nu) \simeq \frac{1}{3}.
\end{align} 
The results in this paper can be scaled to any narrow $Z'$ 
that couples to quarks and muons by the relation:
\begin{align}
|g_{qq}^R| \to |g_{qq}^R|\sqrt{3\times\mathcal{B}(Z'\to\mu^+\mu^-)}.
\end{align}

Search for heavy dilepton resonances by ATLAS~\cite{Aaboud:2017buh} 
and CMS~\cite{CMS:2016abv} set stringent bounds on 
$\sigma(pp\to Z'+ X)\cdot \mathcal{B}(Z'\to\mu^+\mu^-)$,
hence on $g^R_{qq}$ couplings.
The ATLAS result is based on 36 fb$^{-1}$ data, while the CMS result is for 13~fb$^{-1}$.
We use the former~\cite{Aaboud:2017buh}
to extract 95\% credibility level (CL) upper limits on $g^R_{cc}$ and $g^R_{bb}$ couplings,
shown as the purple shaded regions in Fig.~\ref{dyprocess}.
In doing so, we calculate $\sigma(pp\to Z'+ X)$,
where the dominant contribution is from $q\bar q \to Z'$ with subdominant contributions
$qg \to qZ'$ and $gg\to q \bar{q} Z'$ ($q = c$ or $b$), at leading order (LO)
for fixed $m_{Z'}$ and $g^R_{qq}$ by MadGraph5\_aMC@NLO~\cite{Alwall:2014hca} (referred as MadGraph5\_aMC from here on); 
we generate matrix elements (ME) with up to two additional jets in 
the final state\footnote{
We restrict ourselves to up to two additional jets in the final state 
due to computational limitation.} 
with the parton distribution function (PDF) set NN23LO1~\cite{Ball:2013hta},
followed by PYTHIA~6.4~\cite{Sjostrand:2006za} adopting the MLM 
scheme~\cite{Alwall:2007fs} for ME and parton shower (PS) matching and merging.
We, then, rescale the estimated cross section by $|g^R_{qq}|^2$
and extract the upper limit on $|g^R_{qq}|$ for each $m_{Z'}$ 
from the ATLAS result assuming $\mathcal{B}(Z'\to\mu^+\mu^-) \simeq 1/3$.
In Fig.~\ref{dyprocess}, the $5\sigma$ discovery reach\footnote{
Significance is defined by $S/\sqrt{B}$, 
where $S$ and $B$ denote the number of signal 
and background events, respectively.} 
is also given with 3000 fb$^{-1}$ data
for the High Luminosity LHC (HL-LHC).
If the $Z'$ couples to $u$, $d$ or $s$ quark, 
the limits on $g^R_{qq}$ would be much stronger due to a larger PDF ,
i.e. probing a much smaller $g^R_{uu}$, $g^R_{dd}$ or $g^R_{ss}$ coupling 
than that of $g^R_{cc}$ and $g^R_{bb}$.
The details of the cut-based analysis and background 
processes are given in Appendix~\ref{dimuon}.
For sake of a decent $S/B$ ratio, 
we restrict ourselves to $m_{Z'} \lesssim 700$ GeV.

In principle, the methodology in this paper can be applied to left-handed (LH) $qqZ'$ 
couplings $g_{qq}^L$, although there is some subtlety;
that is, the SU(2)$_L$ gauge symmetry relates couplings of
the up- and down-type sector quarks nontrivially.
For instance, a nonzero $g^L_{cc}$ is generally accompanied by a nonzero $g^L_{ss}$
and all possible down-type sector couplings, e.g., $g^L_{dd}$, $g^L_{bb}$ and $g^L_{bs}$,
which are CKM-suppressed.
Hence, one has to deal with multiple couplings simultaneously.
This would complicate the analysis, and we defer to future study.

\section{\boldmath The $cZ'$ process}
\label{czprime}

Having discussed the discovery potential of $ccZ'$ coupling through the DY process, 
we turn to $pp\to c Z' +X \to c \mu^+\mu^- +X$, i.e. the $cZ'$ process, 
which requires tagging of $c$-jet. 
Thanks to recent developments in charm tagging by ATLAS~\cite{atlasctag} 
and CMS~\cite{CMS:2016knj}, it is now possible to study such a process, 
where many phenomenological studies and discussions can already be found~\cite{
Delaunay:2013pja, Perez:2015, Brivio:2015fxa, Hou:2016hgs, 
Iwamoto:2017ytj, Han:2017yhy, Cohen:2017exh, Haisch:2017bpz,Sirunyan:2017pob}. 

\subsection{\boldmath Searching for $cZ'$}
\label{discov}

\begin{table}[t]
\centering
\begin{tabular}{|c|c|c|c|c|c|c|c|}
\hline
    & \, $c$-tagger \,           &    $\epsilon_c$     & $\epsilon_b$
    & $\epsilon_{\mbox{\tiny{light}}}$ \\
\hline
\hline

 \, ATLAS \,    & \ \textbf{Conf1} \   &    0.4                & 0.17            & 0.1    \\
                &\textbf{Conf2}       &    0.2                & 0.1            & \,  0.004 \,   \\ 
 \hline                                                                            
CMS       &\textbf{ctagL}       &   \ 0.9 \            & \ 0.45 \        & \ 0.99 \   \\
           & \ \textbf{ctagM} \      &   \, 0.39 \,            & \, 0.26 \,         & 0.19     \\
                &\textbf{ctagT}       &    0.2               & 0.24            & 0.02   \\
\hline
\end{tabular}
\caption{
ATLAS~\cite{atlasctag} 
and CMS~\cite{CMS:2016knj} 
$c$-, $b$- and light-jet tagging efficiencies $\epsilon_c$, $\epsilon_b$
and $\epsilon_{\mbox{\tiny{light}}}$
for different working points.
}
\label{ctagval}
\end{table}

Let us briefly discuss the present $c$-tagging algorithms.
ATLAS~\cite{atlasctag} gives a range for $b$- and light-jet rejections\footnote{
The mistag rate is defined as the complement of rejection rate.
}
for a fixed value of $c$-tagging efficiency. 
These fixed $c$-tagging efficiencies are presented as curves
(called ``iso-efficiency curve'') in the $b$- vs light-jet rejection plane. 
CMS~\cite{CMS:2016knj} presents similar constant $c$-tagging 
efficiency curves in the $b$- and light-jet mistag efficiency plane. 
For ATLAS iso-efficiency curves, $c$-tagging schemes with high light-jet rejection 
have low $b$-jet rejection rates, and vice versa.
The CMS curves show similar behavior.

\begin{table*}[t!]
\centering
\begin{tabular}{|c|c|c|c|c|c|c|c|c|c|c|c|c|c|}
\hline
$c$-tagger  &&&&&&&&&&\\
WP (ATLAS)             & \ Signal \ & $Z/\gamma^*+c$-jet & $Z/\gamma^*+b$-jet 
                  & $Z/\gamma^*+$ light-jet  & $t\bar{t}$ & $Wt$  & $VV$  & $t\bar t V$    & $tWZ$  &  Total  Bkg.\\                                 
\hline
%
\hline
 \textbf{Conf1}         & 1.34     & 14.52     & 3.04      & 34.66           & 11.52        & 1.11    & 1.37     & 0.01   & 0.01  & 66.24 \\
  \textbf{Conf2}        & 0.67     & 7.26     & 1.79       & 1.39           & 6.77         & 0.65    & 1.61      & 0.01   & 0.001   & 19.48  \\
\hline                          
\end{tabular}
\caption{
 Signal and background cross sections (in fb) after selection cuts for a 150 GeV $Z'$
 (with $g_{cc}^R=0.005$) produced via $pp \to c Z' + X  \to c \mu^+ \mu^-+ X $
 at 14 TeV LHC with ATLAS $c$-tagging schemes, where last column gives 
 total background, with $V$ denoting either $W$ or $Z$ boson.
}
\label{cgzp150atlas}
\end{table*}

\begin{table*}[t!]
\centering
\begin{tabular}{|c|c|c|c|c|c|c|c|c|c|c|c|c|c|}
\hline
$c$-tagger &&&&&&&&&  & \\
WP (CMS)             & \ Signal \  & $Z/\gamma^*+c$-jet & $Z/\gamma^*+b$-jet
       & $Z/\gamma^*+$ light-jet  & $t\bar{t}$ & $Wt$    & $VV$  & $t\bar t V$     & $tWZ$ & Total  Bkg. \\                                 
\hline
%
\hline
 \textbf{ctagL}          & 3.02     & 36.31     & 8.04       & 343.14           & 30.48        & 2.93    & 3.7     & 0.06   & 0.01  & 421.03 \\
%
 \textbf{ctagM}          & 1.31     & 14.16     & 4.64       & 65.85           & 17.61         & 1.69    & 2.1    & 0.02   & 0.001   & 106.08  \\
%
 \textbf{ctagT}          & 0.67     & 7.26      & 4.29       & 6.93           & 16.26         & 1.56    & 1.94    & 0.02   & 0.001   & 38.25 \\
\hline
\end{tabular}
\caption{Same as Table.~\ref{cgzp150atlas}, but for CMS $c$-tagging schemes.}
\label{cgzp150cms}
\end{table*}

The largest background for the $cZ'$ process is $Z/\gamma^* +$ light-jet.
In order to reduce this background, we take 
two $c$-tagging working points (WP) with low light-jet mistag rate 
(i.e. high light-jet rejection) from the ATLAS analysis, 
which we call configuration 1 (Conf1) and configuration 2 (Conf2),
given in the first two rows of Table~\ref{ctagval}.
On the other hand, CMS gives three $c$-tagging WPs called 
$c$-tagger L, M and T (abbreviated as ctagL, ctagM and ctagT in this paper),
which we give in the last three rows of Table~\ref{ctagval}.
For both ATLAS and CMS, WPs with higher $b$-jet rejection could be taken
at the cost of lower light-jet rejection for a fixed $c$-tagging efficiency,
but we do not consider such cases in this study.
Note that these $c$-tagging schemes show mild 
dependence on transverse momentum ($p_T$) and pseudo-rapidity ($\eta$) of the jet. 
For simplicity, we take them to be constant in this study.

To illustrate the discovery potential of the $cZ'$ process, we choose the
benchmark values of mass and coupling
$$
  m_{Z'}=150~\mbox{GeV},~g^R_{cc}=0.005,
$$
setting all other $g^R_{qq}$ couplings in Eq.~\eqref{lepton} to zero.

The $cZ'$ process suffers from several SM backgrounds.
The dominant ones are $Z/\gamma^*+$jet, $t\bar{t}$, $Wt$, 
with smaller contributions from $WW$, $WZ$, $ZZ$, $t\bar t Z$, $t\bar t W$ and $t W Z$. 
There exist non-prompt and fake backgrounds such as 
$W+$jets, QCD multi-jets etc., which we do not consider, 
as these backgrounds are not properly modeled in simulation. 
Due to different tagging efficiencies and mistag rates, 
we separate $Z/\gamma^*+$ jet background into three different categories, 
i.e. $Z/\gamma^* +$ $c$-jet, $b$-jet and light-jet, respectively.

\begin{figure}[b]
\centering
\includegraphics[width=.45 \textwidth]{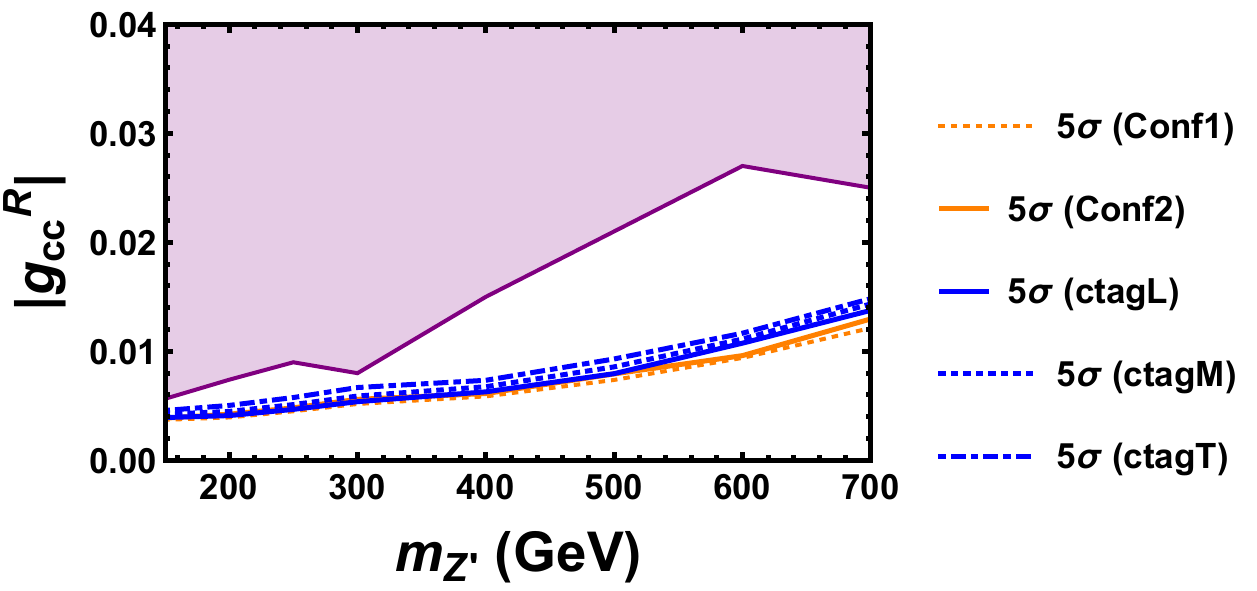}
\caption{
 The $5\sigma$ discovery reach of $pp\to c Z' + X \to c \mu^+\mu^- + X$ 
 process at 14 TeV  with 3000 fb$^{-1}$ data. See text for details.
}
\label{cc_cZp}
\end{figure}

\begin{figure*}[t]
\centering
\includegraphics[width=.3 \textwidth]{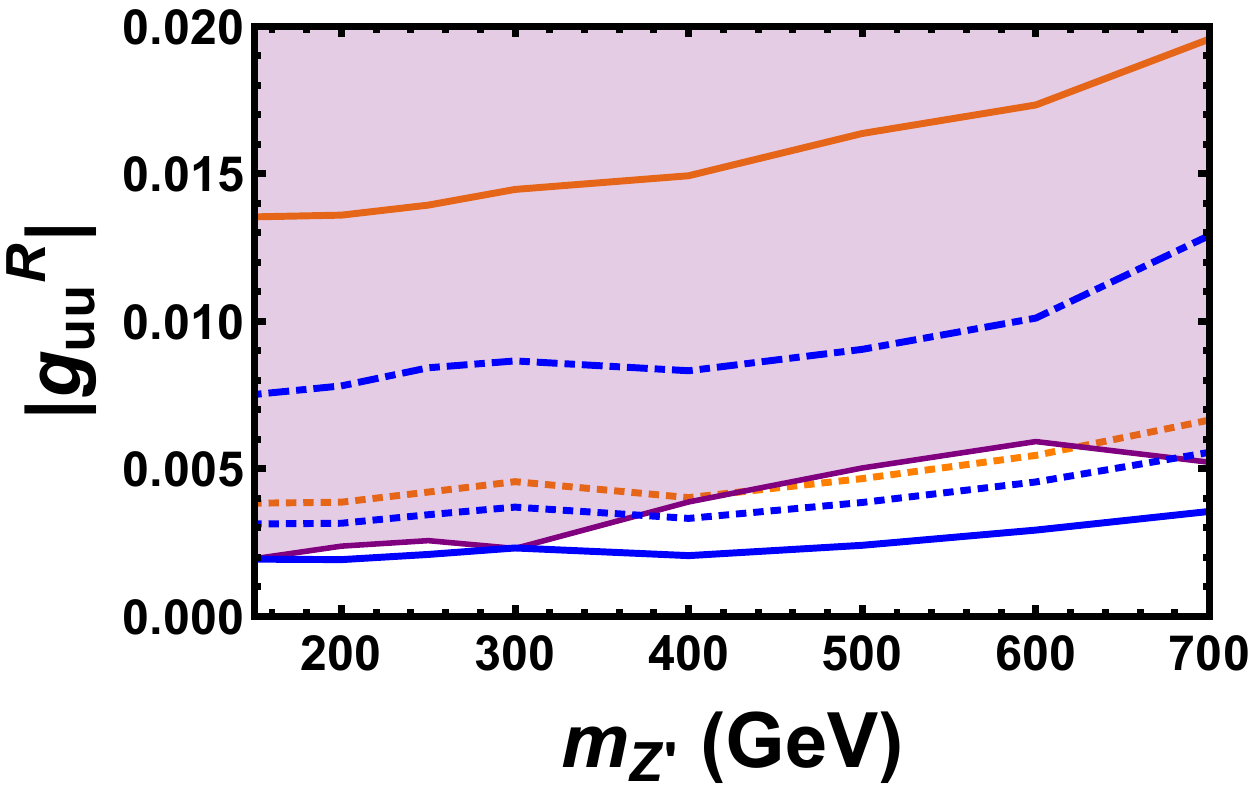}
\includegraphics[width=.3 \textwidth]{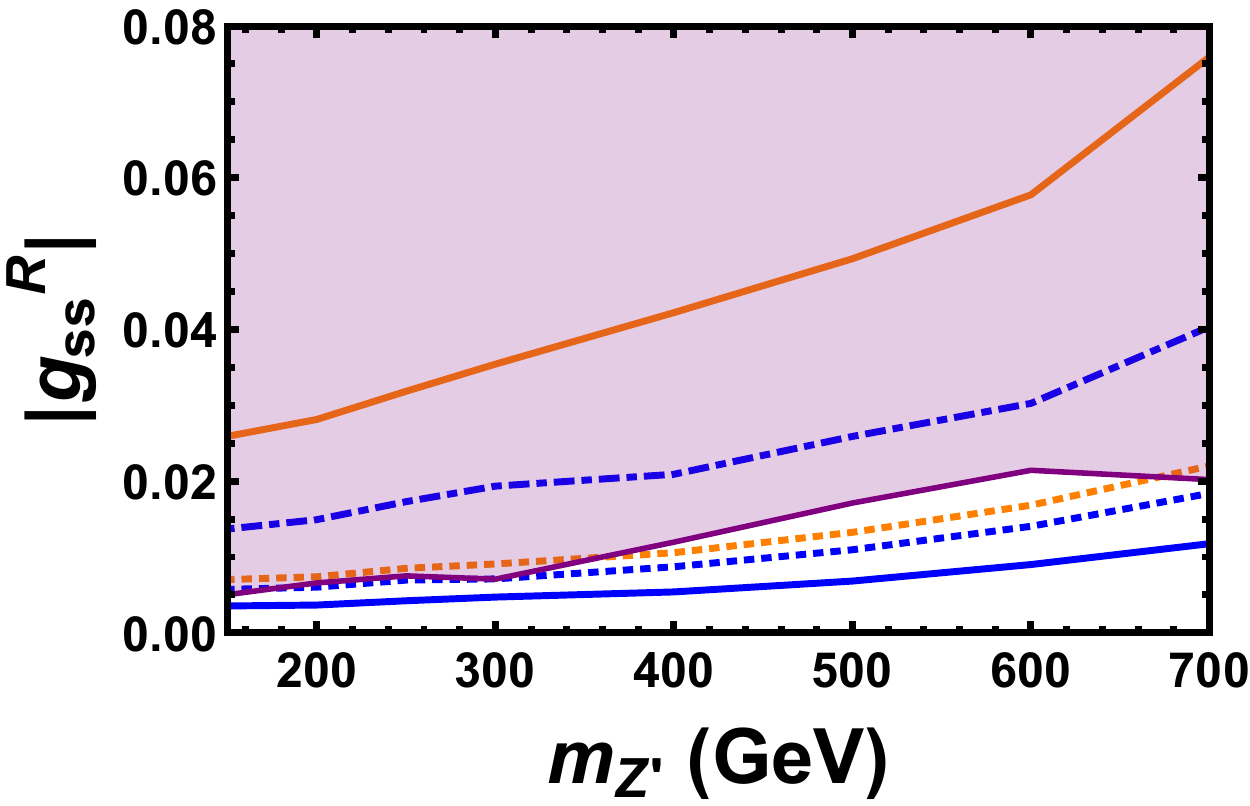}
\includegraphics[width=.3\textwidth]{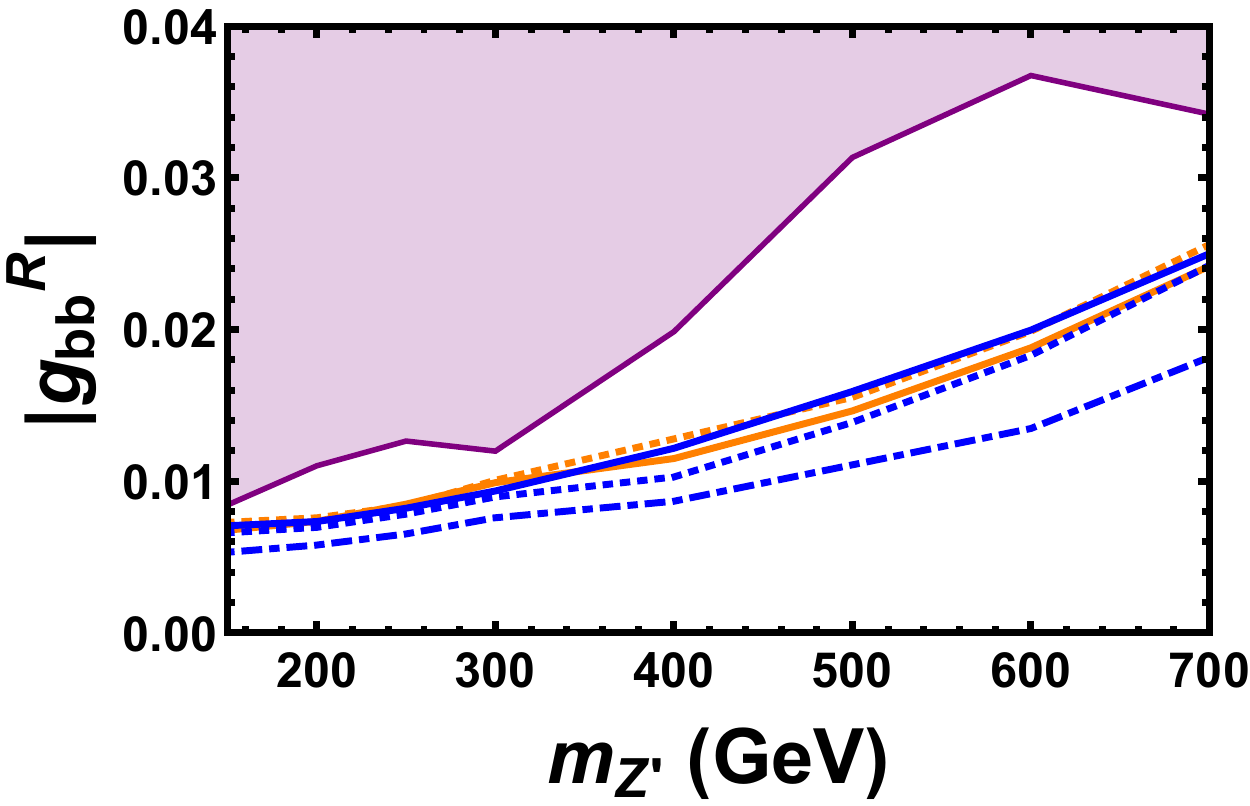}
\caption{
 The $5\sigma$ contours of fake $cZ'$ arising from 
 $g_{qq}^R$ coupling at 14 TeV with 3000 fb$^{-1}$
 (color schemes as in Fig.~\ref{cc_cZp}).
}
\label{fakecc}
\end{figure*}

Signal and background events are generated at LO 
in the $pp$ collision with $\sqrt{s}=14$ TeV via the Monte Carlo event generator
MadGraph5\_aMC@NLO with the PDF set NN23LO1, interfaced to 
PYTHIA~6.4 for showering and hadronization. The event samples are finally fed into the fast detector simulator
Delphes~3.4.0~\cite{deFavereau:2013fsa} for inclusion of (CMS-based) detector effects.
For ME and PS matching and merging we followed MLM matching scheme.
To take higher order corrections into account, 
the LO cross section of $Z\,+$ light-jet is normalized 
by a correction factor $1.83$~\cite{Boughezal:2016isb} up to NNLO.
For simplicity we assume correction factors for  the 
$Z+c$-jet and $Z+b$-jet backgrounds to be same as $Z\,+$ light-jet.
The LO $t\bar{t}$ and $Wt$ cross sections are normalized to the NNLO+NNLL ones
by factors $1.84$~\cite{twiki} and $1.35$~\cite{Kidonakis:2010ux}, respectively. 
Furthermore, the LO cross sections of $WW$, $WZ$ and $ZZ$ backgrounds 
are normalized to the NNLO QCD ones by factors 
$1.98$~\cite{Gehrmann:2014fva}, $2.07$~\cite{Grazzini:2016swo}
and $1.74$~\cite{Cascioli:2014yka}, respectively.
The NLO $K$ factors for the $t \bar t Z$
and $t\bar{t} W^-$ ($t\bar{t} W^+$) backgrounds are assumed to be 1.56~\cite{Campbell:2013yla} and
1.35 (1.27)~\cite{Campbell:2012dh}. 
We do not include $K$ factors for the signal and the $tWZ$ background.


We follow Ref.~\cite{Sirunyan:2017pob} closely in our analysis 
for both signal and background.
We select events with two oppositely charged muons and at least one jet.
Normalized event distributions can be found in Appendix~\ref{dist} for 
transverse momenta of the two muons
and leading $c$-jet,
and the invariant mass of 
a $\mu^+\mu^-$ pair.
We require the leading and subleading muons 
to have $p_T^{\mu_1}> 50$ GeV, $p_T^{\mu_2}> 40$ GeV, respectively.
The transverse momenta of the leading jet in an event should be $p_T^{j} > 45$ GeV.
The minimum separation between two muons ($\Delta R _{\mu\mu}$) and 
the separation between any muon and the leading jet ($\Delta R _{\mu j}$) 
are required to be $> 0.4$.
The maximum pseudo-rapidity ($\left|\eta\right|$) of both muons 
and the leading jet in an event are required to be $< 2.5$.
The jets are reconstructed using anti-$k_T$ algorithm 
with radius parameter $R=0.5$.
To reduce contribution from $t\bar{t}$ and $Wt$ backgrounds,
events with missing transverse energy $(E_T^{\rm miss})>$ 40 GeV are rejected. 
Finally, we impose an invariant-mass cut $|m_{\mu\mu}-m_{Z'}| < 15$ GeV on the 
two oppositely charged muons in an event. 
If an event contains more than one $m_{\mu\mu}$ combination, 
the combination closest to $m_{Z'}$ is selected. 
The impact of the selection cuts on the signal and backgrounds 
are given in Table~\ref{cgzp150atlas} (based on ATLAS $c$-tagging)
and Table~\ref{cgzp150cms} (based on CMS $c$-tagging). 

The ATLAS Conf1 and Conf2 schemes may discover $cZ'$ process 
with $930$ fb$^{-1}$ and $1090$ fb$^{-1}$ integrated luminosities, respectively. 
The dominant background contribution for Conf1 is from 
$Z/\gamma^*+$ light-jet, while $Z/\gamma^*+c$-jet 
constitute the second largest background. 
This is distinctly different for Conf2: $Z/\gamma^*+c$-jet 
and $t\bar t$ provide the dominant and second largest contributions.
A larger $c$-tagging efficiency makes Conf1 superior to Conf2 for discovery. 
Similarly, ctagL, ctagM and ctagT for CMS could discover $cZ'$ process 
with 1150 fb$^{-1}$, 1550 fb$^{-1}$ and 2120 fb$^{-1}$ integrated luminosities. 
The ctagL requires roughly the same luminosity as ATLAS Conf2,
although the $c$-tagging efficiencies and 
$b$- and  light-jet mistag rates are different. 
The larger $c$-tagging efficiency of ctagL is balanced by 
higher mistag rates for light- and $b$-jets. 
The smaller $c$-tagging efficiencies make the $cZ'$ process 
harder to discover for ctagM and ctagT.

Following the same selection cuts,\footnote{
Our study is for illustration, and we do not optimize the selection cuts for each $m_{Z'}$.
We, however, checked a possible impact of such a cut optimization.
The largest impact would be obtained by narrowing the invariant mass window 
$|m_{\mu\mu}-m_{Z'}| < 15$ GeV for a light $Z'$:
we found, for $m_{Z'} = 150$ GeV, the 5 GeV window leads to enhancement 
in the signal significance by $\sim30\%-34\%$, depending on the $c$-tagging scheme.
We found effects of changing the $p_T$ cuts for the muons and leading $c$-jet
are minor, once we impose the $|m_{\mu\mu}-m_{Z'}|$ cut, which tends to select
events with higher $p_T$ muons for a higher $Z'$ mass.
}
we extend our analysis for $Z'$ mass up to 700 GeV.
The discovery reaches for the ATLAS Conf1 (orange dotted), Conf2 (orange solid), 
CMS ctagL (blue dot-dashed), ctagM (blue dotted) and ctagT (blue solid ) 
with 3000 fb$^{-1}$ data are given in Fig.~\ref{cc_cZp}.

\subsection{\boldmath Fake $cZ'$}
\label{light}

Signal for $cZ'$ process could 
arise from light- and $b$-jet mistags, 
which we display in Fig.~\ref{fakecc} for the cases of
$g^R_{uu}$ (left), $g^R_{ss}$ (middle) and $g^R_{bb}$ (right) couplings, 
for LHC at $\sqrt{s}=$ 14 TeV with 3000 fb$^{-1}$ data. 
The purple shaded regions 
correspond to 95\% CL upper limits extracted from Ref.~\cite{Aaboud:2017buh}.
Let us take a closer look.

\begin{table*}[htb]
\centering
\begin{tabular}{|c|c|c|c|c|c|c|c|c|c|c|c|c|c|}
\hline
\ Signal \  & $Z/\gamma^*+b$-jet & $Z/\gamma^*+c$-jet & $Z/\gamma^*+$light-jet & $t\bar{t}$ & $Wt$ &$VV$   & $t\bar t V$ & $tWZ$ & \ Total Bkg.\  \\
            
\hline
%
\hline
              1.31     &  11.35       &  6.89    &  2.53         & 53.21       &  4.28    & 2.69  & 0.05   & 0.01   & 81.01 \\
\hline                          
\end{tabular}
\caption{
 Signal and background cross sections (in fb) after selection cuts 
 for a 150 GeV $Z'$ (with $g_{bb}^R=0.005$)
 via $pp \to b Z' + X  \to b \mu^+ \mu^-+ X$ (plus conjugate process)
 at 14 TeV LHC.
}
\label{mumub_cross}
\end{table*}

The fake  $cZ'$ signals depend on the upper limits 
on $qqZ'$ coupling and the $c$-tagging schemes adopted. 
The extraction of upper limits involves the 
underlying DY process $q\bar q \to Z'$,
which depends on the initial state quark PDFs, 
and is also proportional to $\left|g^R_{qq}\right|^2$. 
On the other hand, fake $cZ'$ signals can originate from
$qg\to q Z'$ and its conjugate process. 
Although also proportional to $\left|g^R_{qq}\right|^2$, 
the cross sections are suppressed by the $2 \to 2$ nature 
compared to the DY process, and depend on gluon and quark PDFs. 
Due to high light-jet rejection rates, two $c$-tagging schemes 
Conf2 and ctagT can fully eliminate fake $cZ'$ from light-jets.
That is, the $5\sigma$ contours for them lie in the excluded regions for both 
$g^R_{uu}$ and $g^R_{ss}$ couplings in the $Z'$ mass range studied,
unlike Conf1, ctagL and ctagM,
which excludes only some $m_{Z'}$ regions.

\begin{figure*}[t]
\centering
\includegraphics[width=.35 \textwidth]{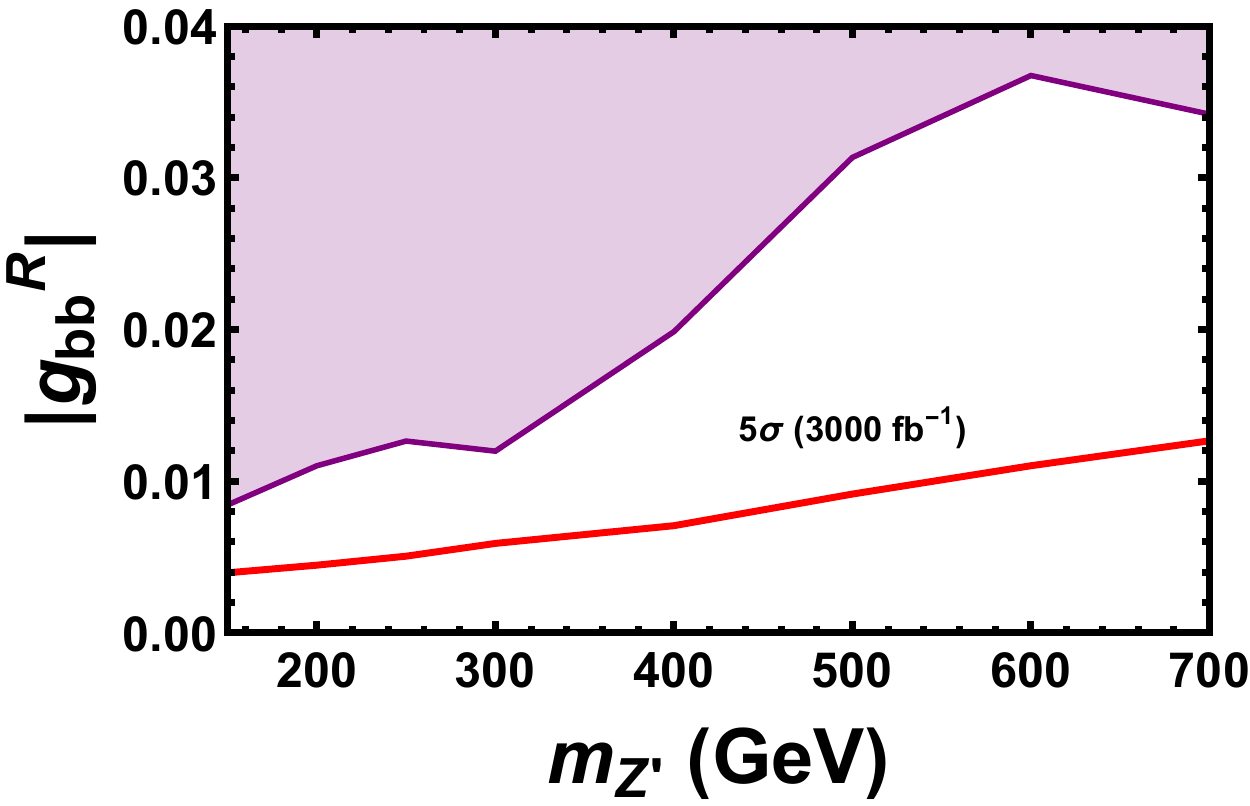}
\includegraphics[width=.35 \textwidth]{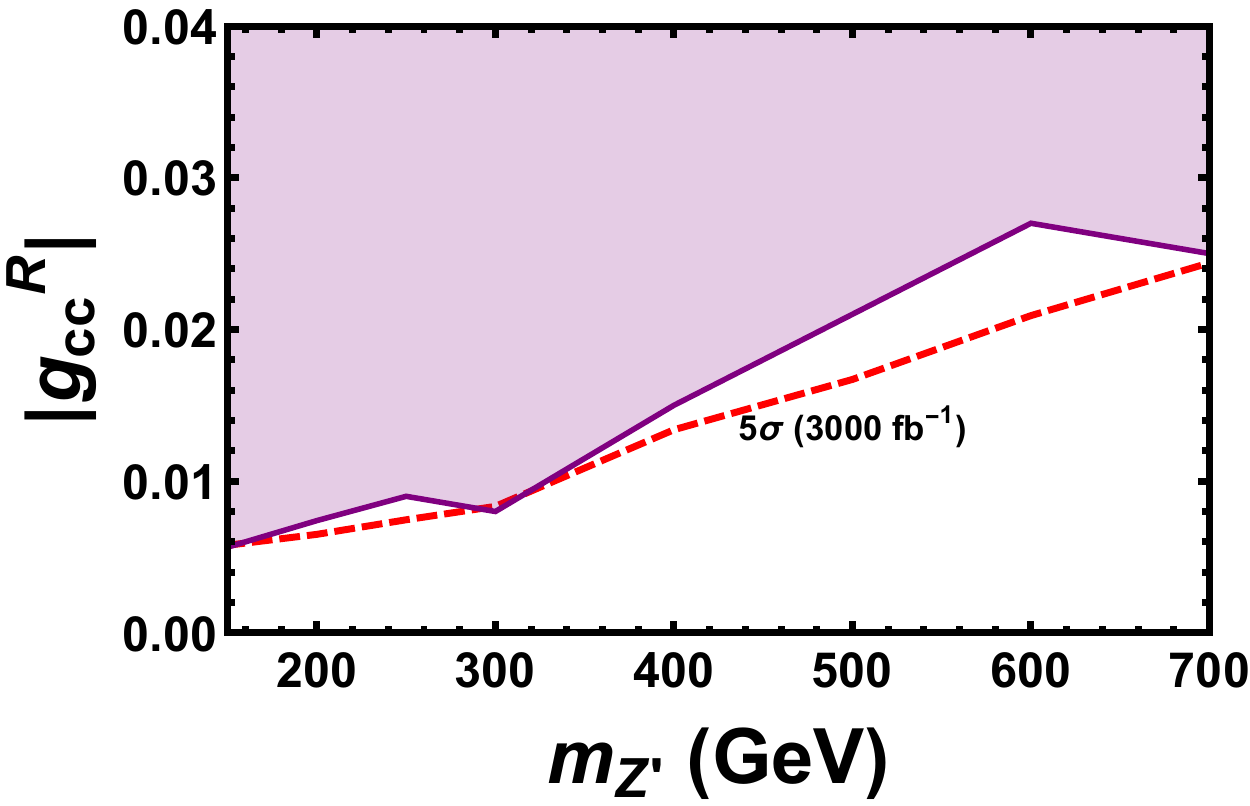}
\caption{
 Discovery reach of $bZ'$ originating from $g^R_{bb}$ (left)
 and $g^R_{cc}$ (right) couplings at 14 TeV LHC with 3000 fb$^{-1}$ data. 
}
\label{mumub}
\end{figure*}

None of these schemes, however, shows promise 
in reducing fakes from $b$-jet misidentification,
since all schemes have considerable $b$-jet mistag rates. 
This can be seen from the rightmost panel of Fig.~\ref{fakecc}.
The high light-jet rejection and low $c$-tagging efficiency (to reduce the 
dominant $Z/\gamma^*+$light- and  $Z/\gamma^*+c$-jet backgrounds) 
make ctagT performing the worst. 
However, although having same $c$-tagging efficiency and even lower light-jet rejection, 
the lower $b$-jet mistag rate of Conf2 makes it perform better than ctagT. 
Our choice of high light-jet, but moderate $b$-jet, rejections 
allows the possibility of fake $cZ'$ arising from $bbZ'$ coupling.
We thus turn to scrutinize this issue in the next section.

\section{\boldmath The $bZ'$ process}
\label{bzprime}

\subsection{\boldmath Searching for $bZ'$}
\label{disbzprime}

If the discovery of DY produced $Z'$ is due to $bbZ'$ coupling, 
it implies $bg\to b Z'\to b \mu^+ \mu^-$ (and its conjugate) 
could also be discovered at the LHC. 
To illustrate the potential for $pp\to b Z' + X \to b \mu^+\mu^- +X$ 
at LHC, we adopt similar strategy as before, 
and take the following benchmark for mass and coupling:
$$m_{Z'}= 150~\mbox{GeV}, \;  g^R_{bb}=0.005.$$

We follow the same cut-based analysis as in previous section,
except the tag jet is now a $b$-jet.
We incorporate in Delphes $p_T$ and $\eta$ dependent  $b$-tagging efficiencies.
The rejection factor of the light-jets are taken as 137~\cite{ATLAS:2014ffa}. 
For simplicity, we assume the correction factors to 
the LO background cross sections generated by MadGraph5\_aMC
to be the same as in previous section, and do not multiply
$K$ factor for the signal. The signal and background cross sections after selection cuts
are given in Table~\ref{mumub_cross}. The required luminosity 
to discover the 150~ GeV $Z'$ is 1180~fb$^{-1}$. 
Our analysis is further extended up to $m_{Z'}= 700$ GeV,
as shown in the left panel of Fig.~\ref{mumub}. 
For simplicity we choose the same selection cut as in $cZ'$ process to generate Fig.~\ref{mumub}.

\subsection{\boldmath Fake $bZ'$}

Mistagged light- or $c$-jets can also produce fake $bZ'$ signals at the LHC, 
but the required $g^R_{qq}$ couplings ($q = u,\, d,\, s$)
to produce fake $bZ'$ at $5\sigma$ with 3000 fb$^{-1}$ are 
already disallowed by heavy resonance DY searches~\cite{Aaboud:2017buh}. 
This attests to the excellent performance of $b$-tagging algorithms 
in reducing light-jet contributions.
However, fake $bZ'$ can still arise from mistagged $c$-jets,
{except two tiny mass windows around $m_{Z'} \sim 150$ and $300$ GeV,
as can be read from the right panel of Fig.~\ref{mumub}
for the $5\sigma$ reach with 3000 fb$^{-1}$.}
We infer that, if no $Z'$ is observed via DY with $\sim 250$ fb$^{-1}$ dataset,
one can rule out the possibility of fake $bZ'$ from the $ccZ'$ coupling at LHC.

Even if a $Z'$ is discovered via DY with $\sim 100 $ fb$^{-1}$ or smaller dataset,
one can still eliminate the possibility of fake $bZ'$ from $ccZ'$ coupling
by combining $bZ'$ and $cZ'$ searches.

For instance, a 600 GeV $Z'$ with $g^R_{cc} = 0.02$, which can be discovered
with 110 fb$^{-1}$ of data via the DY process, requires 1310 fb$^{-1}$ of data
to give fake $bZ'$ signals at 5$\sigma$; however, observing $cZ'$ does not
take long after the discovery of the DY process (e.g., 160 fb$^{-1}$ for
Conf2 and 350 fb$^{-1}$ for ctagT; see the left panel of Fig.~\ref{dyprocess}, Fig.~\ref{cc_cZp},
and  the right panel of Fig.~\ref{mumub}).
In general, fake $bZ'$ from $ccZ'$ coupling, if observed, should be preceded 
by the discovery of $cZ'$ with a smaller dataset.
A similar argument holds for fake $cZ'$ from $bbZ'$ coupling:
after discovery via DY induced by $bbZ'$, fake $cZ'$ can emerge,
but it should be preceded by discovery of $bZ'$
for all five $c$-tagging schemes
(see the right panels of Fig.~\ref{dyprocess} and Fig.~\ref{fakecc}, 
and left panel of Fig.~\ref{mumub}).
Therefore, the simultaneous search for $cZ'$ and $bZ'$ can reveal
if the coupling behind DY production is $ccZ'$ or $bbZ'$.

\section{\boldmath 
 Presence of both $cZ'$ and $bZ'$ Processes}
 \label{bothnonzero}

We have so far studied the discovery potential of $cZ'$ and $bZ'$ processes 
with a nonzero $ccZ'$ or $bbZ'$ coupling exclusively.
However, all $uuZ'$, $ddZ'$, $ssZ'$, $ccZ'$ and $bbZ'$ couplings could in principle coexist. 
If any of the first three couplings involving light quarks are nonzero, 
we might discover $Z'$ in DY process, without subsequent discovery of 
$cZ'$ and/or $bZ'$ processes which can be easily discerned by using 
both $c$- and $b$-tagging algorithms.

A more interesting scenario is when both $ccZ'$ and $bbZ'$ couplings are nonzero,
but all other couplings to light quarks vanish. 
These couplings would give rise to both $cZ'$ and $bZ'$ processes,
depending on their individual strengths.
In order to investigate such a scenario, we take the following benchmark point:
$$m_{Z'} = 150\, {\rm GeV}, \; g^R_{cc} = 0.003, \; g^R_{bb} = 0.005.$$
These $g^R_{cc}$ and $g^R_{bb}$ values remain within respective allowed
regions, as well as $\sigma (pp\to Z'+X)\cdot \mathcal{B} (Z' \to \mu^+\mu^-)$
within 95\% CL upper limit set by ATLAS~\cite{Aaboud:2017buh}.
Larger $g^R_{cc}$ and $g^R_{bb}$ would be in tension with $\sigma\cdot\mathcal{B}$ 
upper limit.

This benchmark can be discovered in the  DY process with just 210 fb$^{-1}$
integrated luminosity, followed by a discovery in the $bZ'$ process
with 870 fb$^{-1}$ data, which is lower than the one quoted 
for case $g^R_{bb} = 0.005$ alone in Sec.~\ref{disbzprime}.
The $cZ'$ process would emerge later, 
at 2370~fb$^{-1}$ (Conf1), 2420~fb$^{-1}$ (Conf2), 2570~fb$^{-1}$
(ctagL), 2600~fb$^{-1}$ (ctagM) or 1740~fb$^{-1}$ (ctagT).
%
The benchmark thus illustrates the possibility of uncovering both 
charm and bottom couplings of a new $Z'$ resonance, 
and the efficacy of the HL-LHC. 
Further sharpening of heavy flavor tagging tools would be helpful.

\section{Summary}
\label{summary}

We analyze the possibility to probe the coupling structure of 
a relatively weakly coupled $Z'$ via the $q g \to qZ'$ process, 
adopting $c$- and $b$-tagging algorithms of ATLAS and CMS at 14 TeV LHC. 
Such a resonance would appear first in the Drell-Yan process. 
Our study shows that, if a $Z'$ is discovered first via 
the $pp\to Z'+ X \to \mu^+\mu^-+X$ DY production, 
one could then discover $cg \to cZ'$ and $bg \to bZ'$ processes at the HL-LHC. 
We illustrate with two different $c$-tagging schemes from ATLAS, 
chosen to optimally reduce $Z+$ light-jet background,
but maintaining moderate $c$-tagging efficiencies. 
We also adopt three $c$-tagging working points from CMS in our analysis.

The $cZ'$ process could arise from misidentification of light- or $b$-jets.
Fake $cZ'$ from light-jet misidentification can be excluded by existing data,
if one adopts ATLAS Conf2 or CMS ctagT scheme.
However, none of the $c$-tagging schemes can rule out the possibility of 
fake $cZ'$ from mistag of $b$-jets.
In order to eliminate fake $cZ'$ from finite $bbZ'$ coupling, 
we advocate simultaneous study of $cZ'$ and $bZ'$ processes. 
We find that a nonzero $bbZ'$ coupling would give genuine $bZ'$ and fake $cZ'$ signatures.
Conversely, a nonzero $ccZ'$ coupling can give genuine $cZ'$ and fake $bZ'$, 
within the allowed region of $ccZ'$ coupling.
The latter possibility can be eliminated in the near future
if no $Z'$ emerges in the DY process with $\sim 250$ fb$^{-1}$ data.
Our study is based on the current status of $c$-tagging algorithms.
Any future improvement in $c$-tagging would only improve the analysis.

It would be interesting if both $ccZ'$ and $bbZ'$ couplings are nonzero.
We illustrate with one such representative scenario, i.e.
for a 150 GeV $Z'$ with $g^R_{cc}=0.003$ and $g^R_{bb}=0.005$. 
We find that 210 fb$^{-1}$ data is needed for DY discovery, 
which would be followed by discovery of the $bZ'$ process with 870~fb$^{-1}$, 
while the $cZ'$ process would emerge much later with integrated luminosities ranging 
from $\sim1740$~fb$^{-1}$ to 2600~fb$^{-1}$, depending on $c$-tagging scheme. 
This scenario differs from cases when either $g^R_{cc}$ or $g^R_{bb}$ vanish. 
For example, when only $g^R_{cc} = 0.005$ is nonzero, 
DY discovery for a 150 GeV $Z'$ would be followed by 
discovery in the $cZ'$ process, without emergence of
subsequent $5\sigma$ signature of fake $bZ'$ process, 
even with full HL-LHC data. 
However, if $g^R_{bb} = 0.005$ is the only nonzero coupling, 
DY process would be followed by discovering the $bZ'$ process. 
The highest attainable fake $cZ'$ signature in this scenario would be about $4.4\sigma$.

We have not included backgrounds associated 
with fake and nonprompt sources, systematic uncertainties and QCD corrections for the signal,
which would induce some uncertainties to our results.
Furthermore, we have not included the uncertainties from scale dependence and PDF with the latter being large for 
the heavy quarks, in particular for $b$ quark. 
The PDF uncertainties for $c$ or $b$
quark initiated processes are discussed in Refs.~\cite{Buza:1996wv,Maltoni:2012pa}, 
while a detailed discussion on PDF choices and their uncertainties for Run 2 of LHC can be found in 
Ref.~\cite{Butterworth:2015oua}. All these effects would impact on the extracted upper limits on 
the $g_{cc}^R$ and $g_{bb}^R$ couplings, as well as our estimated luminosities for discovery.

Our study illustrates that new resonances could still emerge at the LHC,
and large integrated luminosities can probe weaker couplings, or unravel more detail.
Given that our study was partly motivated by flavor ``anomalies''~\cite{Aaij:2013qta,Aaij:2014ora,Aaij:2015oid,Aaij:2017vbb},
associated flavor of $Z'$ production could shed more light 
on potential new physics indications from the flavor sector.
Of course, one would certainly search for other $Z'$ decay modes,
such as $Z' \to \tau^+\tau^-$ implied by Eq.~(1).
\\


\noindent
{\bf Note Added.}
While revising the manuscript, we noticed CMS released a new 
result~\cite{Sirunyan:2018exx} for the dilepton resonance search 
with 36 fb$^{-1}$ data of the 13 TeV LHC.
We checked resulting 95\% CL upper limits on different $g^R_{qq}$ couplings,
with the procedure to interpret the CMS results discussed in Ref.~\cite{Hou:2017ozb},
and found that the new CMS limits~\cite{Sirunyan:2018exx} are comparable 
to the ATLAS limits with 36 fb$^{-1}$ data~\cite{Aaboud:2017buh},
except for $m_{Z'} \sim 500$ GeV, where the CMS gives slightly stronger limits
due to a sharp downward fluctuation in its observed data.
We confirmed that the new CMS limits do not impact on our conclusion.

\begin{figure*}[t]
\centering
\includegraphics[width=.4 \textwidth]{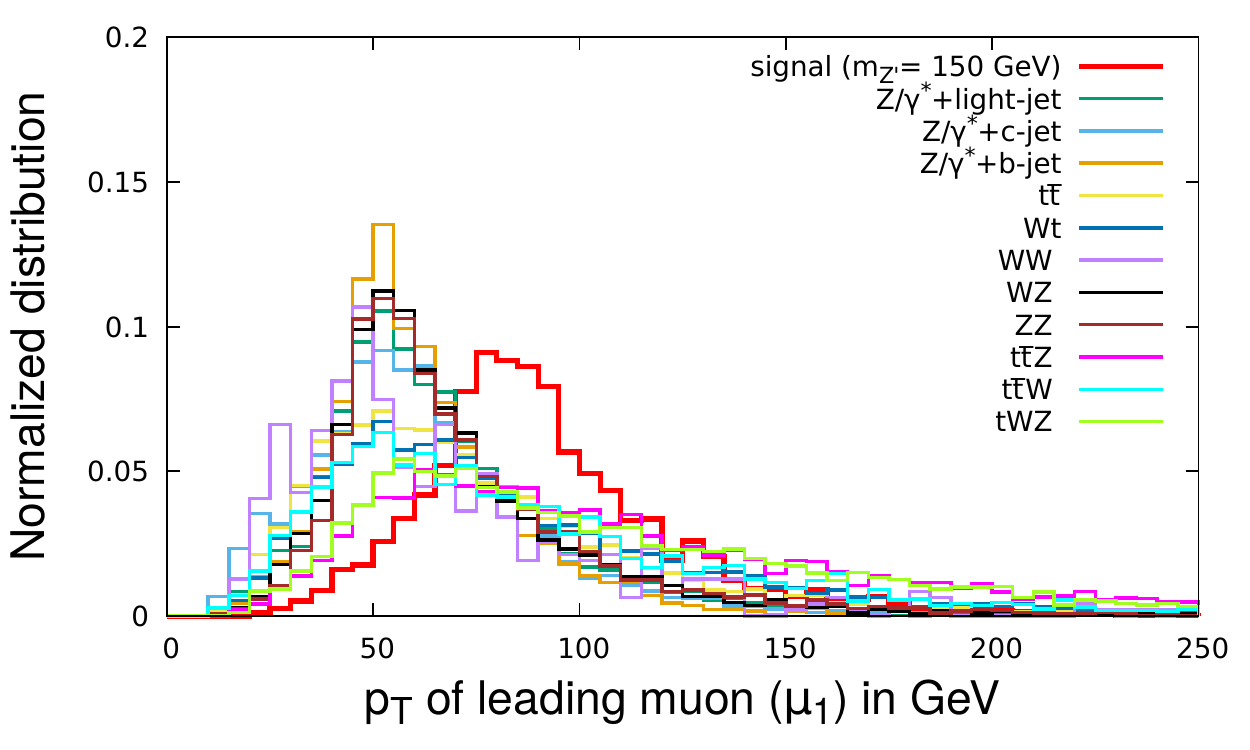}
\includegraphics[width=.4 \textwidth]{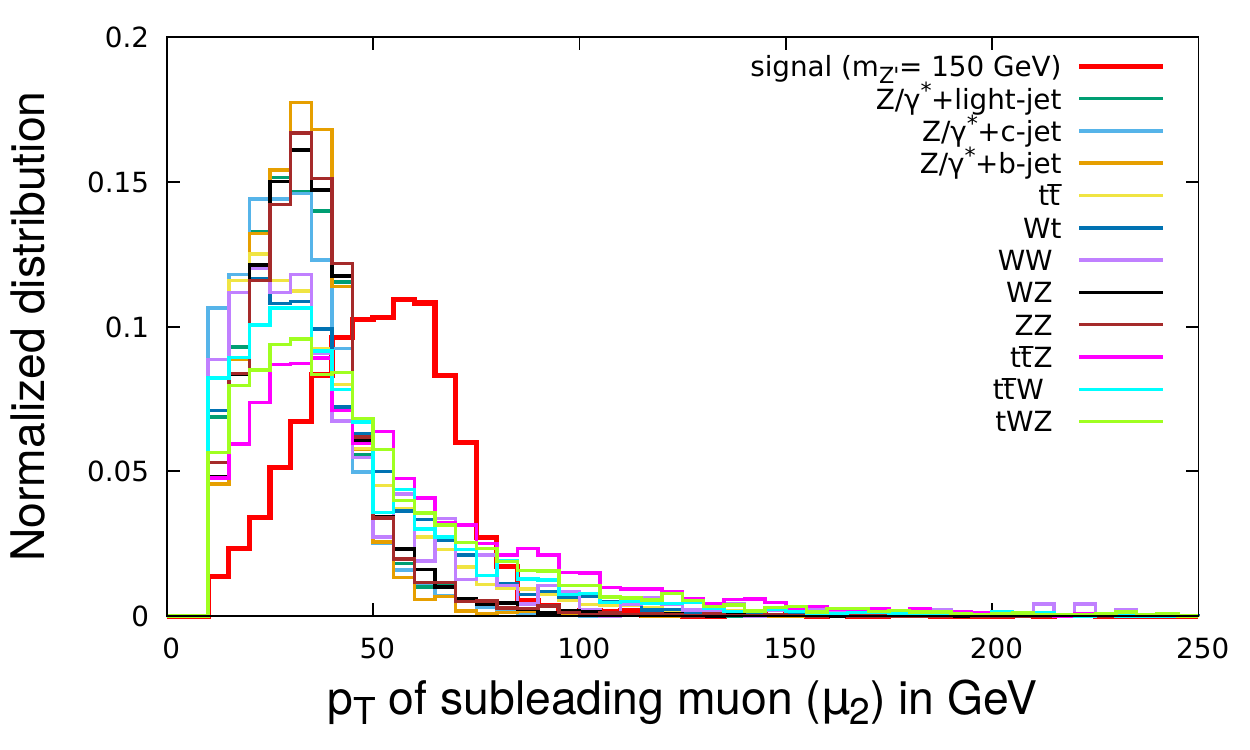}
\includegraphics[width=.4 \textwidth]{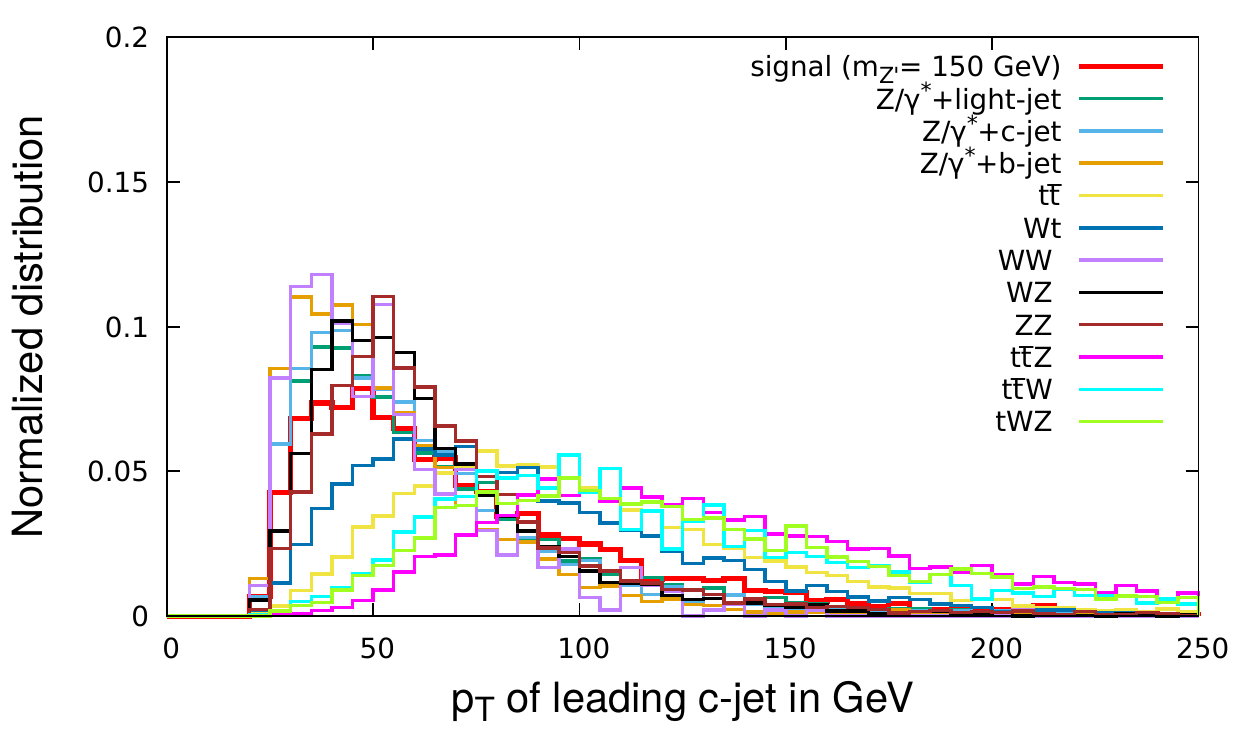}
\includegraphics[width=.4 \textwidth]{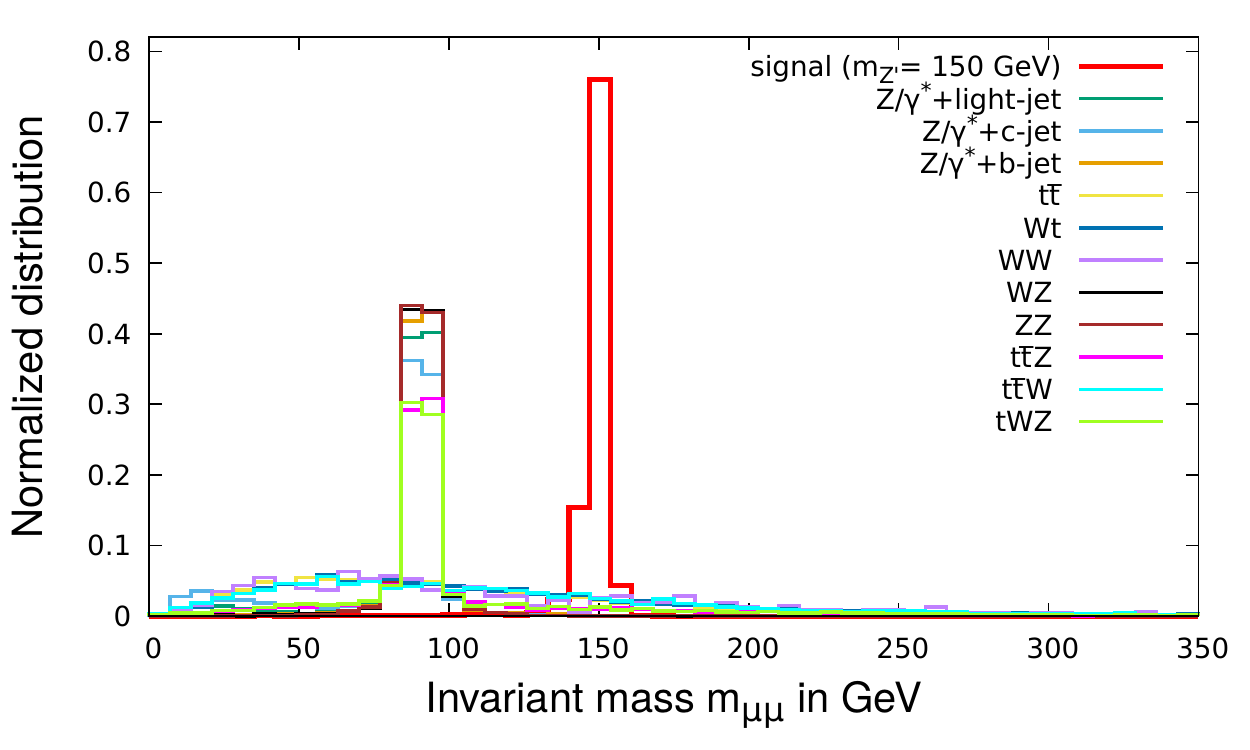}
\caption{
  Normalized distributions of various kinematic variables 
  for the $cZ'$ process ($m_{Z'}=150$ GeV) 
  and its backgrounds: 
  transverse momenta of the leading muon (upper left),
  subleading muon (upper right) and leading $c$-jet (lower left),
  and the dimuon invariant mass (lower right). See text for details.
}
\label{kin}
\end{figure*}

\begin{acknowledgments}
This work is supported by grants NTU-ERP-106R8811 and NTU-ERP-106R104022,
and MOST 105-2112-M-002-018, 106-2811-M-002-187 and 106-2112-M-002-015-MY3.
WSH thanks G.N. Taylor for hospitality.
\end{acknowledgments}

\appendix

\section{Signal and background for DY process}
\label{dimuon}

The dominant backgrounds associated with DY production are 
the $Z$/$\gamma^*$ and $t\bar{t}$ processes, with subdominant 
contributions from $Wt$, $VV$, $t \bar t V$ and $tWZ$ productions, where $V= W,\, Z$.
Selected events should contain two oppositely-charged muons with
transverse momenta $p_T^{\mu}>$ 50 GeV, and an invariant-mass 
cut of $|m_{\mu\mu}-m_{Z'}|<15$ GeV is imposed. 
Signal and background processes are generated at LO via
MadGraph5\_aMC, interfaced to PYTHIA~6.4 and 
fed into fast detector simulator Delphes 3.4.0, following the MLM
prescription for the ME and PS matching and merging. 
The QCD correction factors for the $t\bar t$, $Wt$ and $VV$ 
backgrounds are the same as described in Sec.~\ref{czprime}. 
However, the $Z$/$\gamma^*$ cross section is corrected 
up to NNLO QCD +NLO EW by a factor 1.27,
obtained by FEWZ 3.1~\cite{Li:2012wna}. 
The impact of the selection cuts on different backgrounds 
are given in Table.~\ref{cut_table_zp_150} for various $Z'$ masses. 
Note that, just like $cZ'$ and $bZ'$ processes, 
we do not include $K$ factor for the signal.

\begin{table}[h!]
\centering
\begin{tabular}{|c|ccccccc|c|c|}
\hline
  \ $m_{Z'}$\,(GeV) \        &       150 &     200 &  300    & 400  &500     &600    & 700          \\      
\hline
\hline
%
 Total Bkg.  &  \  2327     &   842    & 177     & 55      &  20     & 9     &  5\\
 \hline
\end{tabular}
\caption{Background cross sections (in fb) for the DY process after
selection cuts, for various $m_{Z'}$ values.
}\label{cut_table_zp_150}
\end{table}

\section{Kinematic Distributions}
\label{dist}

Normalized kinematic distributions for the $cZ'$ process ($m_{Z'}=150$ GeV) 
and its backgrounds are shown in Fig.~\ref{kin}.
Specifically, they are generated with default cuts of MadGraph5\_aMC
for $g^R_{cc} = 0.005$ and ctagT; 
but, other choices for $g^R_{cc}$ and $c$-tagging scheme
should give the same normalized distributions.
The latter is in part because we assume the constant $c$-tagging efficiencies
with respect to $p_T$ and $\eta$ of the jet, but recovering mild dependencies
on them would not affect the normalized distributions significantly.


\end{document}